\documentclass[conference]{IEEEtran}

\usepackage{graphicx}
\usepackage{subfig}
\usepackage{listings}
\usepackage{textcomp}
\usepackage{multirow}
\usepackage{array}
\usepackage{cite}
\usepackage{url}
\usepackage{balance}
\newcolumntype{L}[1]{>{\raggedright\let\newline\\\arraybackslash\hspace{0pt}}m{#1}}
\newcolumntype{C}[1]{>{\centering\let\newline\\\arraybackslash\hspace{0pt}}m{#1}}
\newcolumntype{R}[1]{>{\raggedleft\let\newline\\\arraybackslash\hspace{0pt}}m{#1}}

\graphicspath{{images/}}
\DeclareGraphicsExtensions{.pdf,.jpeg,.png,.jpg,.eps}

\begin{document}
\title{Benchmarking SciDB Data Import on HPC Systems}

\author{\IEEEauthorblockN{Siddharth Samsi,
Laura Brattain,
William Arcand, 
David Bestor, 
Bill Bergeron, 
Chansup Byun, \\
Vijay Gadepally,
Michael Houle, 
Matthew Hubbell, 
Michael Jones,
Anna Klein, 
Peter Michaleas, 
Lauren Milechin, \\
Julie Mullen,
Andrew Prout, 
Antonio Rosa, 
Charles Yee, 
Jeremy Kepner and
Albert Reuther
\IEEEauthorblockA{MIT Lincoln Laboratory\\
Lexington, MA}}}

\maketitle

\begin{abstract}
SciDB is a scalable, computational database management system that uses an array model for data storage. The array data model of SciDB makes it ideally suited for storing and managing large amounts of imaging data. SciDB is designed to support advanced analytics in database, thus reducing the need for extracting data for analysis. It is designed to be massively parallel and can run on commodity hardware in a high performance computing (HPC) environment. In this paper, we present the performance of SciDB using simulated image data. The Dynamic Distributed Dimensional Data Model (D4M) software is used to implement the benchmark on a cluster running the MIT SuperCloud software stack. A peak performance of 2.2M database inserts per second was achieved on a single node of this system. We also show that SciDB and the D4M toolbox provide more efficient ways to access random sub-volumes of massive datasets compared to the traditional approaches of reading volumetric data from individual files. This work describes the D4M and SciDB tools we developed and presents the initial performance results. This performance was achieved by using parallel inserts, a in-database merging of arrays as well as supercomputing techniques, such as distributed arrays and single-program-multiple-data programming. 
\end{abstract}


\IEEEpeerreviewmaketitle

\section{Introduction}
\let\thefootnote\relax\footnotetext{This material is based upon work supported by the Assistant Secretary of Defense for Research and Engineering under Air Force Contract No. FA8721-05-C-0002.  Any opinions, findings and conclusions or recommendations expressed in this material are those of the author(s) and do not necessarily reflect the views of the Assistant Secretary of Defense for Research and Engineering.}SciDB is an open-source database management system that uses an array data model~\cite{stonebreaker2011,brown2010}. The array-based data model provides support for parallel processing, efficient sparse storage, and in-database linear algebra operations that are well suited for the storage and analysis of biomedical imaging data. SciDB is a full ACID (atomicity, consistency, isolation, durability) database management system that guarantees repeatability of results across multiple users operating on the same data. Additionally, it supports array versioning. When using array versioning, SciDB creates new versions of an array instead of modifying an existing array. One of the unique advantages of SciDB is it's ability to perform fast range selects and joins. This capability is achieved by storing data in chunks, in the same order as in the original coordinate system. By storing data in this manner, data that are close to each other can be accessed very quickly by reducing the number of reads necessary to access a given range of data. SciDB also allows a user-settable overlap between chunks of data to speed up applications such as spatial filtering of images for which fewer data reads are necessary at the boundaries to read in the required arrays. 

The array data model of SciDB makes it well suited for managing multidimensional image data. With advances in image acquisition techniques~\cite{tomer,leica}, it is possible to generate multi-Terabytes of high-resolution 2D and 3D images of biological specimens. For example, Tomer et. al. report volumetic image data of 4.8 terabytes generated from a tissue of size 0.5x0.5x0.5 cubic-micrometers. Such large datasets necessitate the development of new approaches to data management, storage and analysis. Open sources tools such as OMERO~\cite{omero} and bisque~\cite{bisque} offer solutions for managing a variety of microscopy image data. These tools are designed to be extensible and can be used from programming languages such as python, MATLAB and Java. However, in our experience, using these tools to manage very large datasets that require out-of-core processing has been challenging because of their attempt to load entire datasets into memory. This can make the system unusable for processing big imaging data. SciDB stores data in it's own native format and provides database functions that allow fast query and extraction of the data of interest. It is also designed to be deployed on massively parallel processing systems and can take advantage of multiple cores on a system. 

The Dynamic Distributed Dimensional Data Model (D4M)~\cite{kepner2012} provides a uniform framework, based on the mathematics of associative arrays~\cite{d4m2.0}, that can be applied to diverse domains such as cyber, bioinformatics, free text, and social media data. D4M can also be used to perform linear algebraic operations inside a database~\cite{vijay2015}. In this paper, we extend D4M to encompass multidimensional arrays in SciDB using image data management as an example application. 

D4M also works seamlessly with the pMatlab (http://www.ll.mit.edu/pMatlab) ~\cite{pmatlab,kepner2009} parallel computing environment which allows high performance parallel applications to be constructed with just a few lines of code. pMatlab uses a single-program-multiple-data (SPMD) parallel programming model and sits on top of a message passing interface (MPI) communication layer. SPMD and MPI are the primary tools used in much of the parallel computing world to achieve the highest levels of performance on the world’s largest systems (see hpcchallenge.org). These tools can also be used for achieving high performance on SciDB.

Section~\ref{sec:technology} describes SciDB, D4M, pMatlab, and the MIT Supercloud system. Section~\ref{sec:design} describes the driving application behind this work and Section~\ref{sec:benchmark} describes the experiments conducted and the measured performance. Finally, Section~\ref{sec:summary} summarizes the results. 

\section{Technologies}
\label{sec:technology}
\begin{figure}[ht]
  \centering
  \includegraphics[width=20pc]{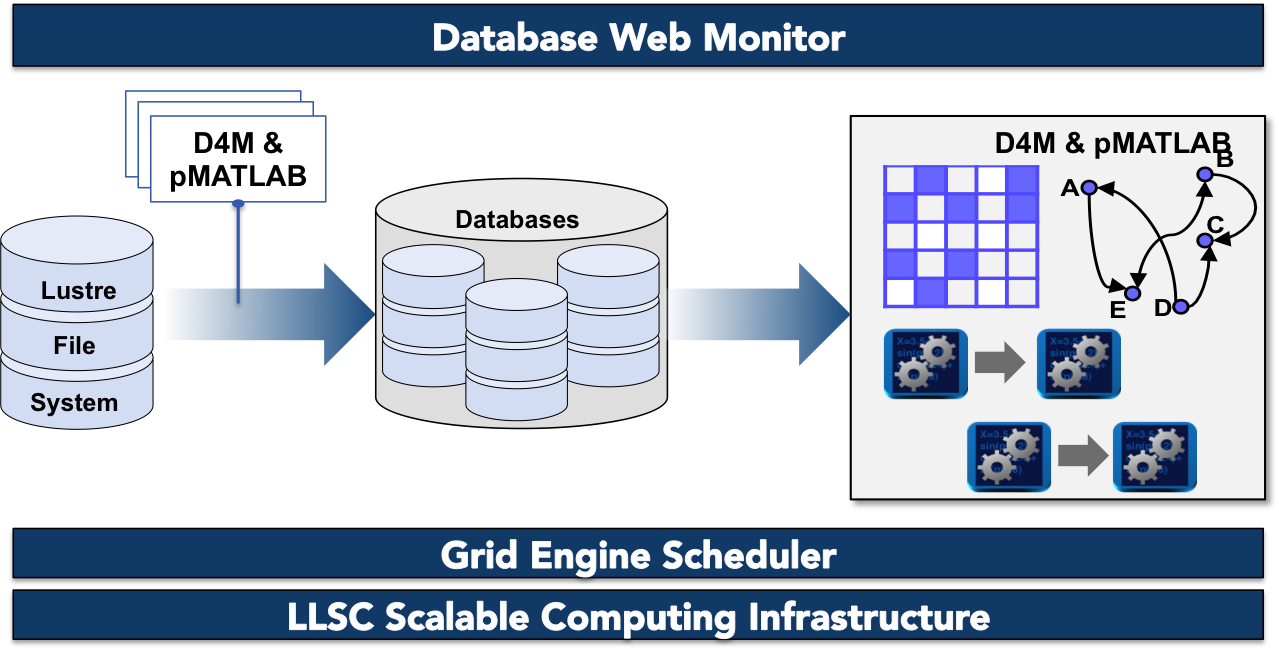}
  \caption{MIT SuperCloud architecture consists of seven components. (1) Lustre parallel file system for high performance file I/O, (2) D4M \& pMatlab ingest processes, (3) parallel databases, (4) D4M \& pMatlab analytic processes, (5) database web monitor page, (6) Grid Engine scheduler for allocating processes to hardware and (7) the TX-Green supercomputer.}
  \label{fig:supercloud}
\end{figure}

A variety of technologies were used to conduct benchmarks for SciDB data ingest. Together, these technologies make up the MIT SuperCloud (see Figure~\ref{fig:supercloud}) and are described in the following subsections. 

\subsection{SciDB}
\label{ssec:scidb}
SciDB is a scalable, computational database management system designed for advanced analytics on multidimensional data~\cite{scidb}. SciDB uses an array data model for storage; thus, it is ideally suited for scientific data such as images, time series data, weather data, and sensor data. Data are stored in the user-defined co-ordinate system such that data that are close to each other in the coordinate system are stored in the same chunk on disk. This storage mechanism has a significant advantage for performing operations such as selecting ranges or joining multiple arrays. Additionally, the number of files read can be minimized because of the ability to specify overlaps in the chunks used to store data on disk. By appropriately specifying the array schema, it possible to optimize the data access and query speeds in SciDB. 

SciDB is built to take advantage of massively parallel processing architectures and is highly scalable on commodity hardware. It also provides the capability to run advanced analytics in database, including parallel linear algebra routines. 

\subsection{D4M analytics library}
\label{ssec:d4m}
D4M is open-source software that provides a convenient mathematical representation of the kinds of data that are routinely stored in spreadsheets and large triple-store databases. Associations between multidimensional entities (tuples) using string keys and string values can be stored in data structures called associative arrays. For example, in two dimensions, a D4M associative array entry might be:

\vspace{6pt}
\centerline{\textbf{A}(\textquotesingle alice \textquotesingle, \textquotesingle bob \textquotesingle) = \textquotesingle cited \textquotesingle \hspace{4pt} or \hspace{4pt} \textbf{A}(\textquotesingle alice \textquotesingle, \textquotesingle bob \textquotesingle) = 47.0}
\vspace{6pt}

The above tuples have a 1-to-1 correspondence with their triple-store representations:

\vspace{6pt}
\centerline{(\textquotesingle alice \textquotesingle,\textquotesingle bob \textquotesingle,\textquotesingle cited \textquotesingle) \hspace{4pt} or \hspace{4pt}  (\textquotesingle alice \textquotesingle,\textquotesingle bob \textquotesingle,47.0)}
\vspace{6pt}

Associative arrays can represent complex relationships in either a sparse matrix or a graph form (see Figure~\ref{fig:d4mfig}). Thus, associative arrays are a natural data structure for performing both matrix and graph algorithms. Such algorithms are the foundation of many complex database operations across a wide range of fields~\cite{kepner2011}. Constructing complex composable query operations can be expressed by using simple array indexing of the associative array keys and values, which themselves return associative arrays:

\begin{figure}[h]
  \includegraphics[width=20pc]{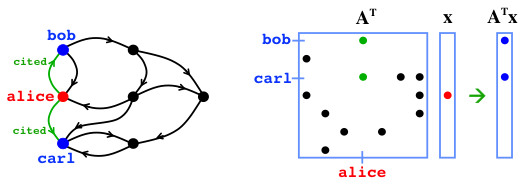}
  \caption{A graph describing the relationship between alice, bob, and carl (left). A sparse associative array A captures the same relationships (right). The fundamental operation of graphs is finding neighbors from a vertex (breadth-first search). The fundamental operation of linear algebra is matrix vector multiply. D4M associative arrays make these two operations identical.  Thus, algorithm developers can simultaneously use both graph theory and linear algebra to exploit complex data.}
  \label{fig:d4mfig}
\end{figure}

\vspace{6pt}
\begin{tabular}{l l}
\centering
\textbf{A}(\textquotesingle alice \textquotesingle,:) & alice row \\
\textbf{A}(\textquotesingle alice bob \textquotesingle,:) & alice and bob rows \\
\textbf{A}(\textquotesingle al* \textquotesingle,:) & rows beginning with al \\
\textbf{A}(\textquotesingle alice : bob \textquotesingle,:) & rows alice to bob \\
\textbf{A}(1:2, :) & first two rows \\
\textbf{A} == 47.0 & subarray with values 47.0
\end{tabular}
\vspace{6pt}

The composability of associative arrays stems from their ability to define fundamental mathematical operations whose results are also associative arrays. Given two associative arrays A and B, the results of all the following operations will also be associative arrays: 

\vspace{6pt}
\begin{tabular}{c c c c c}
\textbf{A} + \textbf{B} \hspace{5pt}& \textbf{A} - \textbf{B} \hspace{5pt}& \textbf{A} \& \textbf{B} \hspace{5pt}& \textbf{A} $\mid$ \textbf{B} \hspace{5pt}& \textbf{A} * \textbf{B} \\
\end{tabular}
\vspace{6pt}

Measurements using D4M indicate these algorithms can be implemented with a tenfold decrease in coding effort when compared to standard approaches~\cite{kepner2012}.

\subsection{pMatlab parallel computing library}
\label{ssec:pmatlab}
pMatlab is open-source software that allows a Matlab program (mathworks.com) or a GNU Octave program (octave.org) to be launched in parallel.  In a pMatlab program all \textit{Np} parallel instances of the program persist for the life of the program, have a unique process identifier (PID), and can directly communicate with all the other instances of the programs. The communication between each PID is handled by message passing. In addition, pMatlab provides scalable mechanisms for creating distributed arrays so that each PID knows exactly which part of the array it owns and where to find all the other pieces.

pMatlab implements the distributed arrays parallel programming model to achieve high performance on the largest computers in the world.  This model gives the application precise control of its computations and communications when it is running on a parallel computing system. 

\subsection{Lustre parallel file system}
\label{ssec:lustre}
The MIT SuperCloud has two forms of storage: distributed and central.  Distributed storage is comprised of the local storage on each of the compute nodes and this storage is used for running Hadoop and database applications.  Central storage is implemented using the open-source Lustre parallel file system (lustre.org) on a commercial storage array.  Lustre provides high performance data access to all the compute nodes, while maintaining the appearance of a single filesystem to the user.  The Lustre filesystem is used in most of the largest supercomputers in the world.

The MIT SuperCloud leverages both types of storage to dynamically start, stop, checkpoint, relocate, and restart (or clone) databases by storing their data in the Lustre filesystem when the databases are stopped.  This dynamic database management system allows many more SciDB databases to be hosted on the system than would otherwise be possible.  Groups of users can quickly create their own databases to share data among themselves without interfering with other groups.  In addition, because all the database instances are running directly on the compute nodes, they can run at maximum performance.

\subsection{Grid Engine scheduler}
\label{ssec:sge}
Supercomputers require efficient mechanisms for rapidly identifying available computing resources, allocating those resources to programs, and launching the programs on the allocated resources.  The open-source Grid Engine software (gridscheduler.sourceforge.net) provides these services and is independent of programming language (C, Fortran, Java, Matlab, etc.) or parallel  programming model (message passing, distributed arrays, threads, map/reduce, etc.).
The Grid Engine scheduler coordinates the starting and stopping of SciDB database instances in the MIT SuperCloud. SciDB users authenticate by using a web page that shows them only the databases they are allowed to access.  They can then start and stop any of these databases.  When a database is started, Grid Engine determines the computing requirements of the database, finds the computing resources, allocates them to the database, copies all the database files to the appropriate computing nodes, assigns dynamic alias domain name entries to the compute nodes, and starts the database processes.

\subsection{TX-Green hardware}
\label{ssec:txgreen}
The TX-Green supercomputer consists of 270 HP servers connected to a single, non-blocking 10 GigE Voltaire core switch.  The Lustre central storage system uses a 1 Petabyte Data Direct Networks (DDN) storage array and a 5.0 Petabyte Seagate storage array that are directly connected to the core switch. This architecture provides high bandwidth to all the nodes and the central storage. Each compute node has 32 cores (x86 instruction set), 128 Gigabytes of memory, and 12 Terabytes of storage.  The storage on each compute node is hot-swappable RAID5 so that each node can tolerate one drive failure.

TX-Green is housed in an HP EcoPOD mobile data center that uses ambient air cooling to maximize energy efficiency.  The EcoPOD is located near a hydroelectric dam that delivers clean energy that does not contribute greenhouse gases to the environment.

The MIT SuperCloud software stack, which contains all the systems and applications software, resides on every node.  Hosting the application software on each node accelerates the launch of large applications (such as SciDB) and minimizes their dependency on the central storage.

\section{Benchmark Design}
\label{sec:design}
SciDB uses a multidimensional array data model for storage. This model makes it well suited for data such as images (2D or 3D), sensor data, and time-series data. Our benchmark is motivated by a biomedical imaging application that involves 3D volumetric data. Advances in imaging technologies are enabling the acquisition of larger and higher resolution biomedical datasets. Imaging techniques such as CLARITY ~\cite{clarity2013}  have the potential to significantly advance our understanding of brain function by enabling molecular and optical interrogation of brain tissue. However, a significant challenge in this area is the management and analysis of 3D volumetric data generated by using such techniques. Several commercial~\cite{imaris,metamorph} and open source tools~\cite{imagej,omero,vaa3d} are available for the analysis and visualization of biomedical imaging data. Each tool has its advantages and disadvantages, but a common limitation in many of these systems is the inability to analyze and/or visualize datasets that are significantly larger than the total amount of memory available on the system. Additionally, many of these packages are focused on single-client visualization and may not have a programmatic way to serve image data to a variety of clients. 

Traditional parallel computing approaches to the analysis of large image datasets involves the use of multiple processors to analyze subsets of images, with each parallel process reading a subset of images files in order to access the data of interest. SciDB in conjunction with pMatlab enables a new mode of parallel operation where a stack of 2D images can be treated as a true volume. SciDB gives us the ability to efficiently extract random, multi-dimensional data using appropriate queries. Consider a 3D array in SciDB with the following schema: 

\lstset{basicstyle=\ttfamily\footnotesize,breaklines=true}
\begin{lstlisting}
  vol3d<val:uint8> [row=1:4096,4096,0,col=1:4096,4096,0,slice=1:1000,1,0]
\end{lstlisting}

This schema defines a 3D array in SciDB of size 4096x4096x1000 pixels used to store 8 bit unsigned integer values. By using the SciDB AFL query language, a sub-volume can be extracted as shown below:

\begin{lstlisting}
  between(vol3d, 100, 100, 10, 300, 500, 100);
\end{lstlisting}

This query will extract all values in the cube bounded by rows = 100:300, cols = 100:500 and slice = 10:100. This is a powerful capability that we have extended to D4M to extract sub-volumes from SciDB using standard MATLAB indexing syntax as shown in Listing~\ref{matlab}. 

\lstset{basicstyle=\ttfamily\footnotesize,language=Matlab,breaklines=true,label=matlab}
\begin{lstlisting}[frame=single,caption={D4M example to extract sub-volume}]
DB = DBsetupSciDB('txg-testdb');
T = DB('vol3d<gray:uint8>row=1:4096,4096,0,col=1:4096,4096,0,slice=1:1000,1,0]');
v = T(100:300, 100:500, 10:100);
\end{lstlisting}

\subsection{SciDB data ingest}
\label{ssec:ingest}
\begin{figure}[ht]
  \includegraphics[width=20pc]{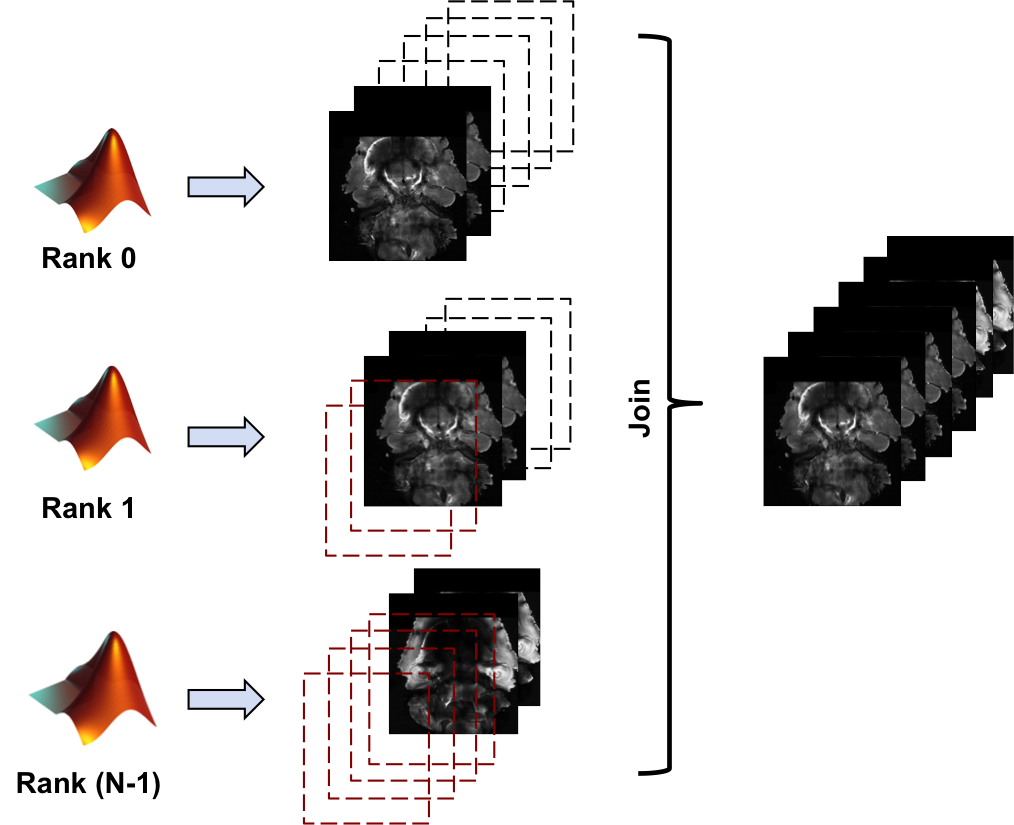}
  \caption{Data is imported into SciDB in a two step process: Step 1 is to ingest each image into a new array and Step 2 is to merge the arrays into a single multi-dimensional array.}
  \label{fig:ingest}
\end{figure}

Since SciDB is a fully ACID database, only one client can ingest data to a given array at a time. In order to maximize the amount of data being ingested, we use multiple MATLAB processes running in parallel. The ACID nature of SciDB results in a serialization of the data ingest process. The solution to this serialization was to ingest data into two stages, as shown in Figure~\ref{fig:ingest}. We launch multiple parallel MATLAB processes using pMatlab. In the first stage, each MATLAB process uses D4M to ingest data into a new array with the appropriate schema. Once all the MATLAB processes have completed their data ingest, the MATLAB process with rank 0 issues a \texttt{merge} statement to combine all the individual arrays into the desired multidimensional array. The merging of individual arrays into a large multi-dimensional array is very fast in SciDB and does not incur any appreciable overhead. 

Data are ingested into SciDB using D4M using a simple ``putTriple'' command as shown in Listing~\ref{ingest}. In this example, slice number 15 of a 1000 slice 3D volume is ingested into an array named \texttt{vol3d}.  

\lstset{basicstyle=\ttfamily\footnotesize,language=Matlab,breaklines=true,label=ingest,caption={Image data ingest using D4M}}
\begin{lstlisting}[frame=single,caption={D4M example to extract sub-volume}]
DB = DBsetupSciDB('txg-testdb');
T = DB('vol3d<gray:uint8>row=1:4096,4096,0,col=1:4096,4096,0,slice=1:1000,1,0]');
im = imread('test-image.tif');
[nr, nc] = size(im);
[rowids, colids] = ind2sub([nr nc], [1:nr*nc]');
slicenum = 15*ones(size(ir));
T = putTriple(T, [rowids colids slicenum], im(:));
\end{lstlisting}

\subsection{Benchmarking data ingest}
\label{sec:benchmark}
\begin{figure}
  \centering
    \subfloat[Ingest rate with single node SciDB instance. Maximum ingest rate: 2.23 million entries/sec.]{
      \includegraphics[width=17pc]{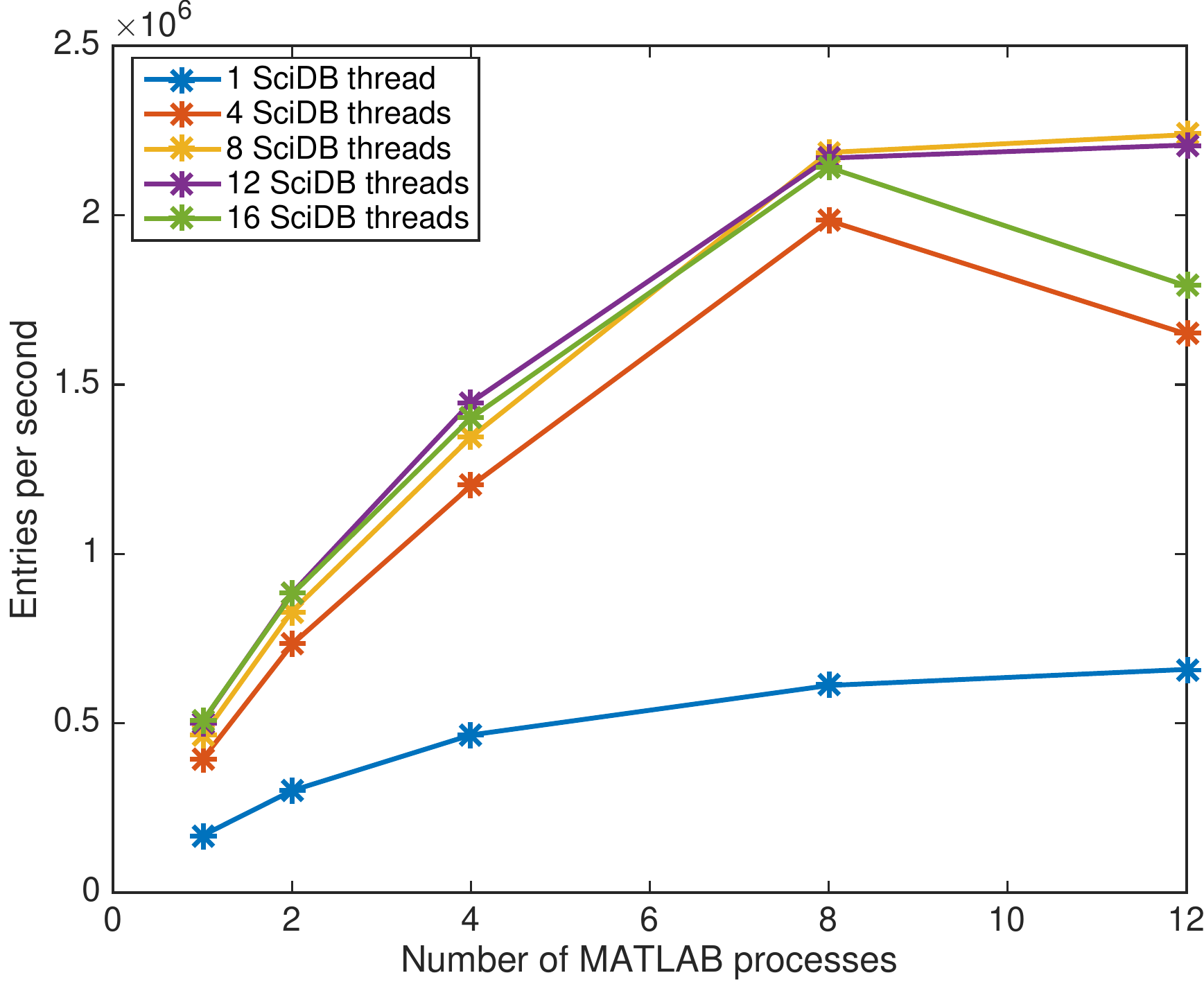}
      \label{subfig:single-node}
    } \\
    \subfloat[Ingest rate with two node SciDB instance. Maximum ingest rate: 2.876 million entries/sec.]{
      \includegraphics[width=17pc]{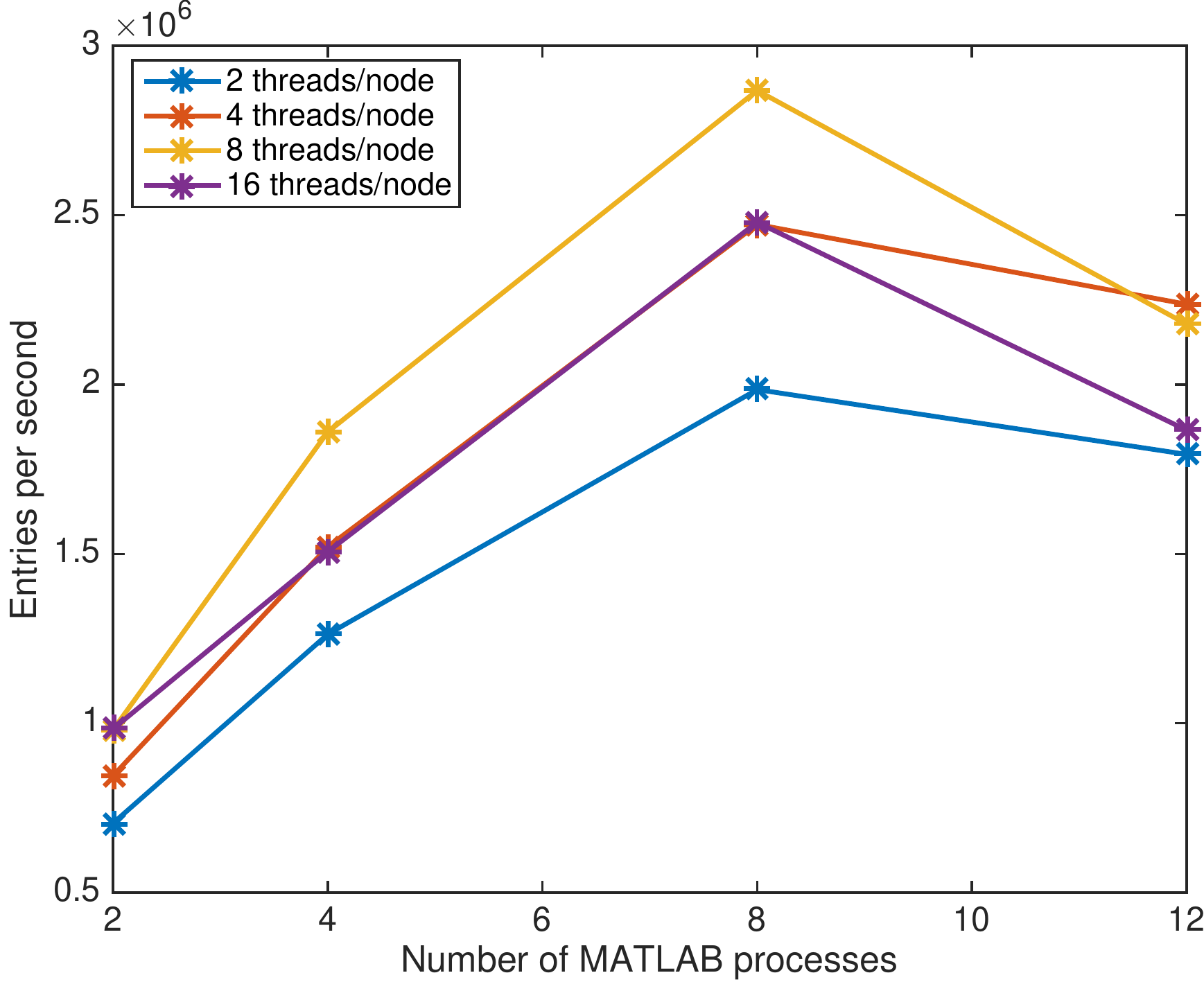}
      \label{subfig:two-node}
    }
    \caption{Timing results for data ingest in SciDB: Data were ingested using multiple parallel processes running on the same compute node as the database. Parallelization was achieved using pMATLAB~\cite{pmatlab}.}
    \label{fig:results}
\end{figure}

For this test, we used randomly generated imaging data to simulate a volume of size 5120 x 5120 x 1000 pixels. The data were ingested into SciDB instances configured with different hardware and software specifications. For a single node instance of SciDB, configurations with 1, 4, 8, 12, and 16 SciDB worker processes were used. A two-node SciDB configuration using 2, 4, 8, and 16 SciDB worker threads per node was also used for testing. Data were imported using 2, 4, 8, and 12 processes running in parallel on the same node as the database. Network bandwidth limitations were minimized by running the ingest processes on the same node as the database. Data import used the SciDB shim interface from MATLAB and D4M. Figure~\ref{fig:results} shows the ingest rates achieved. When importing data into a single-node instance of SciDB, we achieved a maximum ingest rate of 2.23 million entries/second by using 8 parallel MATLAB processes as shown in Figure~\ref{subfig:single-node}. A maximum ingest rate of 2.876 million entries/second was observed when using 8 parallel MATLAB processes importing into a two node SciDB instance. In each of these configurations, the use of more than 8 parallel processes for importing data resulted in a degradation of performance. Similarly, SciDB configurations using more than 8 SciDB instances per node did not result in an appreciable increase in ingest rates and actually resulted in a slower ingest in the case of the two-node SciDB instance, as shown in Figure~\ref{subfig:two-node}.

\section{Summary}
\label{sec:summary}
SciDB is an open-source, array-based database management system designed for advanced analytics on multidimensional data. This paper described the performance of SciDB for ingesting image data. A peak performance of 2,867,910 inserts per second was achieved with a two-node SciDB instance. This performance was achieved using parallel clients running D4M for inserting data into unique arrays and a final merge step. D4M and SciDB offer a new approach to the storage and analysis of large biomedical imaging datasets. To our knowledge, this is the highest database insert rate ever achieved and surpassed our prior single-node record of 800,000 inserts per second achieved with the Accumulo database. By leveraging D4M, it is possible to access large volumetric imaging data stored in SciDB using high-level languages such as MATLAB. The ability to efficiently access any 3D sub-volume from such a dataset gives us a capability that is not easily available with traditional approaches to image data management. Future work in this area includes the implementation of image analysis routines as linear algebra operations that can be run directly in SciDB, thus removing the need to retrieve data from the database. Other areas of research include the development of a more efficient method to import images into SciDB by exploiting sparsity and statistical distribution of the data. 

\balance

\section{Acknowledgment}
The authors would like to thank Prof. Kwanghun Chung (The Picower Institute for Learning and Memory, MIT) for providing imaging data used in the development of D4M extensions and Ms. Dorothy S. Ryan (MIT Lincoln Laboratory) for assistance with proof-reading this article. We would also like to thank David Martinez (Associate Division Head, Cyber Security and Information Sciences, MIT Lincoln Laboratory) and Dr. Jeffrey Palmer (Group Leader, Bioengineering Systems \& Technologies, MIT Lincoln Laboratory) for their support.

\bibliographystyle{unsrt}
\bibliography{IEEEabrv,references.bib}

\end{document}